\documentclass[a4paper, 11pt] {article}
\usepackage{amsfonts,amsmath}
\usepackage{amsthm}

\usepackage{graphicx}
\usepackage{algorithm}
\usepackage[noend]{algpseudocode}
\usepackage{upgreek}

\numberwithin{equation}{section}

\newcommand{\bsym}{\boldsymbol}

\title{A Novel Method of Marginalisation using Low Discrepancy Sequences for Integrated Nested Laplace Approximations}
\author{Paul T. Brown, Chaitanya Joshi, Stephen Joe, and H{\aa}vard Rue}
\begin{document}
\maketitle

\begin{abstract}
Recently, it has been shown that approximations to marginal posterior distributions obtained using a low discrepancy sequence (LDS) can outperform standard grid-based methods with respect to both accuracy and computational efficiency. This recent method, which we will refer to as LDS--StM, can also produce good approximations to multimodal posteriors. However, implementation of LDS--StM into integrated nested Laplace approximations (INLA), a methodology in which grid-based methods are used, is challenging. Motivated by this problem, we propose modifications to LDS--StM that improves the approximations and make it compatible with INLA, without sacrificing computational speed. We also present two examples to demonstrate that LDS--StM with modifications can outperform INLA's own grid approximation with respect to speed and accuracy. We also demonstrate the flexibility of the new approach for the approximation of multimodal marginals.
\end{abstract}

%\begin{keyword}
%Bayesian inference, marginal approximations, integrated nested Laplace approximations, low discrepancy sequences.
%%\MSC{62F15, 65D30}
%\end{keyword}

%\end{frontmatter}

\section{Introduction}
Integrated nested Laplace approximations (INLA) \cite{rue}, is a methodology developed specifically for fast approximate Bayesian inference of latent Gaussian models (LGM) \cite{faramir}, a large class of hierarchical Bayesian models. INLA was developed as a more computationally efficient alternative to more widely used Markov chain Monte Carlo (MCMC) methods. Initially, INLA used grid-based methods of marginalisation to compute marginal hyperparameter posterior distributions \cite{martino, rue}. Generally, a set of points is laid out in a grid structure over a relevant integration region. Along with numerical integration techniques and an interpolant, posterior marginals are constructed. Grid-based methods have been used in different applications, with several variants used, see \cite{hewitt, joshiPHD, ormerod, wilson}. Grids are known to be computationally efficient only in very low dimensions, since the number of points increases exponentially with dimension. \\ 

Computational advancements within INLA have given rise to new algorithms that improve computational efficiency. A set of central composite design (CCD) points \cite{box} is laid out over the hyperparameter space, and marginals are computed using an asymmetric Gaussian interpolant. The CCD strategy requires far less points than grid-based strategies (see \cite{sanchez} for details of how to compute these points). The numerical integration-free method (NIFA) \cite{thiago}, a method bypassing numerical integration altogether, has achieved further computational gains. However, the CCD and NIFA strategies do have the drawback of only allowing for unimodal approximations. For details on both of these strategies, see \cite{thiago}. Recently, a new method called Bayesian inference using sparse grid quadrature evaluation (BISQuE) \cite{hewitt} uses sparse grids (see \cite{novak}) to explore the hyperparameter space and estimate marginal posteriors. BISQuE is similar to INLA as a Bayesian inference framework for hierarchical models, though not necessarily just for latent Gaussian models. \\

An alternative strategy for hyperparameter exploration and estimation, developed by the authors of \cite{joshiArticle}, show that a very general grid-based method can be improved upon by using a low discrepancy sequence (LDS) in the place of a grid. Using a least squares polynomial as an interpolant, \cite{joshiArticle} proved that this LDS-based method of marginalisation (hereafter, referred to as the standard LDS method, or LDS--StM) improved the computational efficiency by reducing the number of points needed, whilst also showing convergence to the true marginal. LDS--StM also allowed for the approximation of multimodal posterior densities.\\

Although LDS--StM performed better than grid-based methods in a general setting, the method itself was not easy to implement within INLA. Issues arose when choosing the degree of the polynomial. If a higher degree was needed (such as with the approximation of skewed, or multimodal distributions), the resulting matrices were sometimes ill-conditioned, and therefore some computation was inaccurate. They also encountered issues with Runge's phenomenon \cite{runge}, where oscillations would form in the tails of the approximations. This paper seeks to expand on the development of the LDS--StM method by applying important modifications to the algorithm to make it more useful in a practical setting. The goal is to fully incorporate a modified LDS-based method into INLA, specifically, for the accurate and efficient estimation of hyperparameter posterior marginal distributions of an LGM. We present the modifications, and two challenging examples of estimating hyperparameter posterior marginals to demonstrate how this modified method works, and how it performs relative to INLA's current methods (INLA's grid and NIFA). The methods presented may also be used in BISQuE as well, though this will require further investigation.\\

This paper is organised as follows. First, we give some necessary background on QMC integration, INLA, and grid-based methods in Section 2. Section 3 gives an overview of the LDS--StM algorithm, and the modifications for implementation in INLA. We provide two examples of the performance of the new method in Section 4, before giving our concluding remarks in Section 5. \\

\section{Background}
\subsection{QMC Integration Rules}
In many instances, we wish to compute an integral
\begin{equation} \label{eq::Integral}
I_f = \int_D f(\bsym{x}) d\bsym{x}, 
\end{equation}
where $\bsym{x}=(x_1, \ldots, x_s)$ is $s$-dimensional, and $D$ is the region of integration, assumed to be the unit hypercube $[0,1)^s$. A typical approximation to $I_f$ in \eqref{eq::Integral} is given by
\begin{equation} \label{eq::MCEstimator}
I_f \approx \dfrac{1}{N} \times \sum_{i=1}^{N} f(\bsym{x}_{(i)}).
\end{equation}
where the points $\bsym{x}_{(i)}, i=1,\ldots,N$ are sampled in the region of integration. One can use a Monte Carlo rule, where the points $\bsym{x}_{(i)}, i=1,\ldots,N$ are sampled randomly, or a QMC rule, where the points are generated deterministically. LDS are a large class of sequences that are generally used in \eqref{eq::MCEstimator} when using a QMC rule. LDS are sequences that hold the property that, for specific values of $N$, its finite subsequence $x_1, \ldots, x_N$ has \emph{low discrepancy} with respect to the Lebesgue measure on the unit hypercube. See \cite{dickBook, lecuyerLem, nieder} for different examples of low discrepancy sequences and definitions of discrepancy. It can be shown via discrepancy theory and the Koksma-Hlawka inequality \cite{lemieux} that for functions with bounded variation (in the sense of Hardy and Krause), the estimator in \eqref{eq::MCEstimator} converges to the true value faster if the points are generated using an LDS, rather than randomly sampled. This result also extends to low discrepancy point sets \cite{pill}, such as the Korobov lattice which we will introduce shortly. For the purposes of this paper, we use the term LDS to refer to both low discrepancy sequences, and low discrepancy point sets. \\

We define here the two deterministic point sets that we use (see Figure 1). First, we define an $s$-dimensional $n$-point grid by
\begin{equation} \label{eq:grid}
{\cal{G}}_{n,s} = \left\{ \left(\frac{g_1}{n}, \cdots, \frac{g_s}{n} \right), g_j = 0,\ldots, n-1, j = 1,\ldots, s \right\},
\end{equation}
where $n$ is the number of unique abscissa points in each dimension, $s$ is the dimension of the point set and $N=n^s$ is the total number of points. These point sets are deterministic but do not have low \emph{discrepancy}. As such, the convergence rate is slow as the number of points increase exponentially as dimension increases \cite{nieder}. \\

The next deterministic point set we define is the Korobov lattice which, unlike the grid, is an LDS. The $s$-dimensional Korobov lattice \cite{joe} with $N$ points is given by
\begin{equation} \label{eq:koroPS}
{\cal{K}}_{N,s,\alpha} = \left\{ \frac{i-1}{N}(1, \alpha, \alpha^{2}, \ldots, \alpha^{s-1}) \; \mbox{mod} \, 1, i = 1, \ldots, N \right\},
\end{equation}
where $\alpha$ is called the generating constant, and is typically chosen to be an integer between $1$ and $N-1$, and relatively prime with $N$. We use the software LatticeBuilder \cite{lattBuild}, to find ``good" generating constants with respect to a discrepancy measure, for fixed $s$, $N$, and constant weights. LatticeBuilder also finds generating constants for extensible Korobov lattices. Extensible lattices have the property that the number of points $N$ can increase whilst retaining existing points and maintaining low discrepancy (see \cite{hickernell} for details). We denote an extensible Korobov lattice with number of points $N$, dimension $s$, and generating constant $\alpha$, by ${\cal{K}}^{*}_{N, s, \alpha}$. \\

\begin{figure}[h]
\begin{center}
\includegraphics[scale = 0.2]{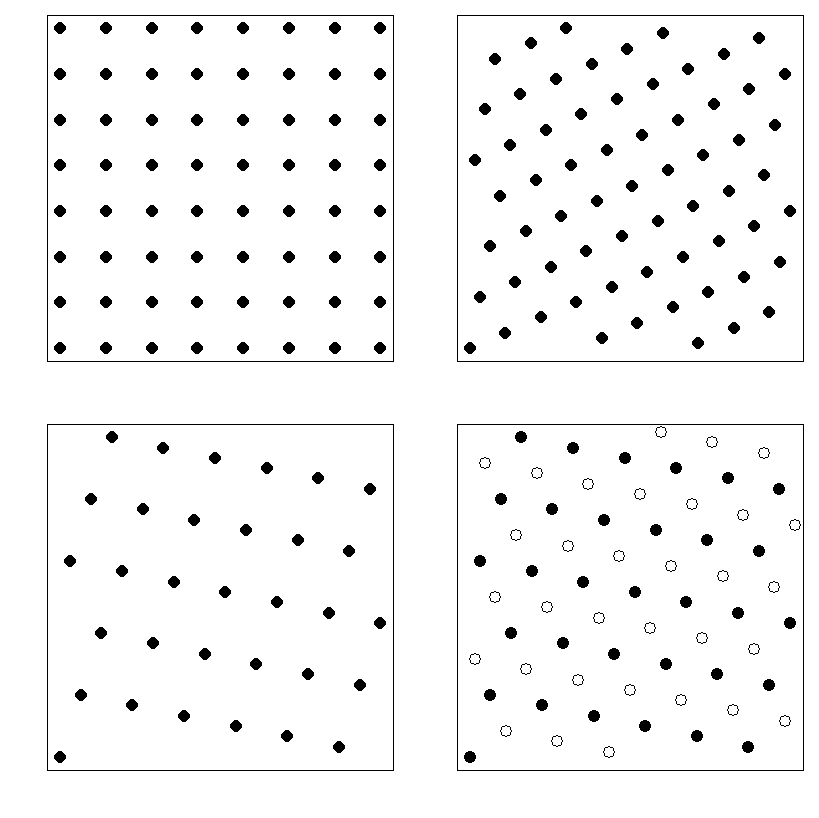}
\caption[grid vs Korobov]{\footnotesize{Point sets: An $8$-point grid ${\cal{G}}_{8,2}$ (top left), a Korobov lattice ${\cal{K}}_{64,2,37}$ (top right), an extensible Korobov lattice ${\cal{K}}^{*}_{32,2,19}$ (bottom left), and an extensible Korobov lattice ${\cal{K}}^{*}_{64,2,19}$ (bottom right) based off the previous lattice. The extra 32 points are shown as hollow circles.}}
\end{center}
\end{figure}

High dimensional integrals arise naturally in the field of Bayesian inference, where marginal posterior distributions are computed from a joint posterior distribution. These integrals are also used in other statistical applications such as the computation of marginal likelihoods. In general, we wish to estimate the following integral
\begin{equation*} \label{eq::Integral2}
f_k(u_k) = \int_{\chi} f(\bsym{u})d\bsym{u}_{-k}, \hspace{3mm} k=1,\ldots,s,
\end{equation*}
where $f_k(u_k)$ is the $k^{th}$ marginal function of the variable $u_k$, and $\bsym{u}_{-k}$ is the variable $\bsym{u}$ excluding the $k^{th}$ component. Taking $\chi$ to be some rectangular region $[\bsym{a}, \bsym{b})$ where $\bsym{a} = (a_1, \ldots, a_s)$ and $\bsym{b}=(b_1, \ldots, b_s)$, we generate a point set $\bsym{x}_{(i)}, i=1,\ldots,N$ in $[0,1)^s$ and use a linear transformation, 
\begin{equation} \label{eq::transformps}
\bsym{u} = \bsym{a} + (\bsym{b}-\bsym{a})\bsym{x},
\end{equation}
so that $\bsym{u}_{(i)}, i=1,\ldots,N$ are scaled to the region $[\bsym{a}, \bsym{b})$. The marginal function $f_k(u_k)$ can be approximated by
\begin{equation} \label{eq::s-1Approximation}
f_k(u_k) \approx \dfrac{\prod_{j=1}^{s} (b_j - a_j)}{b_k - a_k} \times \dfrac{1}{N} \sum_{i=1}^{N} f(u_{1,(i)}, \ldots, u_k, \ldots, u_{s,(i)}),
\end{equation}
where $f(u_{1,(i)}, \ldots, u_k, \ldots, u_{s,(i)})$ are the function evaluations keeping the variable $u_k$ fixed. In the context of Bayesian inference, MCMC algorithms are used to approximate marginal posteriors. However, for some types of hierarchical models such as LGMs, these sampling methods may not necessarily be computationally efficient \cite{rue}. \\

\subsection{INLA}
INLA is a computational Bayesian inference framework that was proposed as a computationally efficient alternative to MCMC methods \cite{rue}. It is particularly designed for LGMs, a widely used class of Bayesian hierarchical model, encompassing basic regression and time series models to complex spatio-temporal point process models and many others (see \cite{rueReview} for a comprehensive list of applications). \\

Given a data set with $n_d$ observations, $\bsym{y}=(y_1, \ldots, y_{n_d})$, assume that the response variable $y_i$ follows a distribution from the exponential family, with its conditional mean $\mu_i$ associated with $\eta_i$ through a link function $\ell$, such that $\eta_i = \ell(\mu_i)$. The LGM has the following additive structure
\begin{equation*}
\eta_i = \beta_0 + \sum_{j=1}^{n_{\beta}} \beta_j u_{ij} + \sum_{k=1}^{n_f} g_k(v_{ik}),
\end{equation*}
where $\beta_0$ is the intercept, $\beta_j$ are the fixed effects for covariates $\bsym{u}$, and $g_k$ are the model components, associated with the covariates $\bsym{v}$. These models can be expressed in a generic three-stage hierarchical formulation. For the first stage, we assume the data $\bsym{y}$ are independent of each other, given  $\bsym{\eta}$ and hyperparameters $\bsym{\theta}_1$, so
\begin{equation*}
 \bsym{y}|\bsym{\eta},\bsym{\theta}_1  \sim  \prod_{i=1}^{n_d} \pi(y_i|\eta_i, \bsym{\theta}_1).
\end{equation*}
The second stage specifies $\bsym{\eta}$ to be a latent Gaussian field with the density function
\begin{equation*}
\bsym{\eta}|\bsym{\theta}_2  \sim  {\cal{N}}(\bsym{0}, \bsym{Q}^{-1}(\bsym{\theta}_2)),
\end{equation*}
where ${\cal{N}}$ is a multivariate Gaussian distribution. In many applications, the latent Gaussian field has conditional independence properties, translating to a multivariate Gaussian distribution with sparse precision matrix $\bsym{Q}(\bsym{\theta}_2)$, otherwise known as a Gaussian Markov random field (GMRF) \cite{ruebook}. The final stage is the hyper-prior stage
\begin{equation*}
\bsym{\theta} = (\bsym{\theta}_1, \bsym{\theta}_2) \sim \pi(\bsym{\theta}),
\end{equation*}
where $\bsym{\theta}$ has $s$ components. \\

INLA works by approximating the posterior density of hyperparameters $\bsym{\theta}|\bsym{y}$ using a Laplace approximation \cite{tierney}. A Gaussian approximation $\tilde{\pi}_G(\bsym{\eta}|\bsym{\theta},\bsym{y})$ for the posterior for the latent parameters is made, and evaluated at the posterior mode $\bsym{\eta}^{*}(\bsym{\theta}) = \underset{\bsym{\eta}}{\operatorname{argmax}}\hspace{1mm}\pi(\bsym{\eta}|\bsym{\theta},\bsym{y})$. Let $\tilde{\pi}(\bsym{\theta}|\bsym{y})$ be the approximation to the joint hyperparameter density. Then the Laplace approximation is given by
\begin{equation} \label{eq::LaplaceApprox}
\left. \tilde{\pi}(\bsym{\theta}|\bsym{y})\propto \dfrac{\pi(\bsym{\theta},\bsym{\eta}|\bsym{y})}{\tilde{\pi}_G(\bsym{\eta}|\bsym{\theta},\bsym{y})} \right|_{\bsym{\eta}=\bsym{\eta}^{*}(\bsym{\theta})}.
\end{equation}
The marginals of the hyperparameters $\theta_k, k=1,\ldots,s$, are found by exploring and then taking specific points in the hyperparameter space, either in a grid structure or otherwise, and using numerical integration and an interpolant \cite{thiago, rue}. Those points are reused in the approximation of the latent parameter marginals $\pi_j(\eta_j|\bsym{y}), j=1,\ldots,n_d$ via the following equation
\begin{equation} \label{eq::latentMarginals}
\pi_j(\eta_j|\bsym{y}) \approx \sum_{i=1}^{N} \tilde{\pi}(\eta_j|\bsym{\theta}_{(i)},\bsym{y})\tilde{\pi}(\bsym{\theta}_{(i)}|\bsym{y})\Delta_i,
\end{equation}
where $\Delta_i$ are the integration weights, $\tilde{\pi}$ is a Laplace approximation of $\pi$, and $\bsym{\theta}_{(i)}, i = 1,\ldots,N$ are the points generated by the integration rule. \\

The number of latent parameters INLA can handle is very large, in the order of $10^5$ \cite{rueReview}. However, to ensure INLA's computational efficiency, it requires that the number of hyperparameters must be small, since those points are re-used to approximate each latent marginal \cite{thiago, rue}. INLA loses much of its speed using a grid strategy for models with five or more hyperparameters, as the number of points increases exponentially with $s$. \\

During the exploration of the hyperparameter space, an optimisation algorithm is used to approximate the mode and Hessian. One way to approximate marginals would be to fit a multivariate Gaussian by matching the mode and curvature at the mode via the Hessian. However, the marginals are often skewed so this approximation is often not accurate. Instead a transformation of the hyperparameter space is performed, which is orthogonal and standardised (see \cite{rue} for details). Let $\hat{\bsym{\theta}}$ denote the transformed space. In $\hat{\bsym{\theta}}$, define
\begin{equation*}
\tilde{\pi}(\hat{\bsym{\theta}}|\bsym{y}) \propto \prod_{k=1}^{s} \tilde{\pi}_k(\hat{\theta}_k|\bsym{y}),
\end{equation*}
where $\tilde{\pi}_k(\hat{\theta}_k|\bsym{y})$ is tangent to the marginal $\tilde{\pi}_k(\theta_k|\mathbf{y})$. To capture the asymmetry of $\tilde{\pi}(\hat{\theta}_{k}|\bsym{y})$, INLA assumes the following structure
\begin{equation} \label{eq::NIFA}
\tilde{\pi}_k(\hat{\theta}_{k}|\mathbf{y}) \propto 
\begin{cases}
\exp\left(-\dfrac{1}{2\sigma_{k+}^2}(\hat{\theta}_{k}^2 - \mu_k) \right), \hat{\theta}_{k} > 0 \\
\exp\left(-\dfrac{1}{2\sigma_{k-}^2}(\hat{\theta}_{k}^2 - \mu_k) \right), \hat{\theta}_{k} \leq 0, \\
\end{cases}
\end{equation}
where the variances $\sigma_{k+}^2$ and $\sigma_{k-}^2$ are approximated assuming each marginal is ``half-Gaussian", with a different variance either side of the mode. The NIFA method assumes the structure in \eqref{eq::NIFA} and then computes $\sigma_{k+}, \sigma_{k-}, k=1,\ldots,s$ without numerical integration (see \cite{thiago} for details). Whilst these methods greatly reduce the number of points, the grid is more accurate overall, since the approximations from the grid do not assume any kind of unimodal structure.

\subsection{Grid-Based Marginalisation Methods}
Grid-based methods are relatively simple in concept and straightforward to implement. A set of grid points is placed in an appropriate integration region. After evaluating the function at those points, we average out over rows and columns and fit an interpolant between the function evaluation means. \\

Let $\pi(\bsym{\theta})$ be any $s$-dimensional density that we wish to marginalise. We drop the dependence on $\bsym{y}$ for notational convenience. Also, let ${\cal{G}}_{n,s} = \{\bsym{g}_{(1)}, \ldots, \bsym{g}_{(N)}\}$ be an equally spaced, $s$-dimensional $n$-point grid, as described in \eqref{eq:grid}. Note that the grid is generally constructed over the unit hypercube, so a scaling transformation, given by
\begin{equation} \label{eq::trans}
T(\bsym{g}) = \bsym{a} + (\bsym{b} - \bsym{a})\bsym{g},
\end{equation}
where $\bsym{g}$ is a grid point, $\bsym{a}$ and $\bsym{b}$ are the lower and upper vertices of the integration region, must be performed on the grid points. We denote this transformed point set by $\{\bsym{\theta}_{(1)}, \ldots, \bsym{\theta}_{(N)}\}$. We present a general method for grid-based marginalisation in Algorithm 1. \\

\begin{algorithm}
\caption{General grid-based method} \label{alg:grid}
\begin{algorithmic}
\State 1) Optimise $\pi(\bsym{\theta})$ for mode $\pi^{*}$ and Hessian $H_{\pi}$
\State 2) Construct an appropriate integration region $[\bsym{a}, \bsym{b})$
\State  3) Generate grid points ${\cal{G}}_{N,s}$ and transform points to $\{\bsym{\theta}_{(j)}\}_{j=1}^{N}$ over $[\bsym{a}, \bsym{b})$ \\ \hspace{3mm} and evaluate $\pi(\bsym{\theta}_{(j)})$ for all $j = 1,\ldots,N$
\For {$k = 1, \ldots, s$}
\State 4) Orthogonally project points $(\bsym{\theta}_{(j)}, \pi(\bsym{\theta}_{(j)}))$ for $j=1,\ldots,N$ onto \\ \hspace{9mm} $[a_k, b_k)$
\State 5) For $l=1,\ldots,n$, where $n$ is the number of abscissa points, compute \\ \hspace{9mm} pointwise means
\begin{equation} \label{eq::PWM}
\hat{\pi}(\theta_{k,l}) = \dfrac{1}{n^{s-1}} \sum_{i_1=1}^{n} \cdots \sum_{i_{s-1}=1}^{n} \pi(\theta_{1,(i_1)}, \ldots, \theta_{k,l}, \ldots, \theta_{s, (i_{s-1})})
\end{equation}
\State 6) Fit interpolant through pointwise means and normalise for $\tilde{\pi}_k(\theta_k)$
\EndFor
\end{algorithmic}
\end{algorithm}

Step 1 can be achieved through any quasi-Newton algorithm, and this is used to construct an integration region $[\bsym{a}, \bsym{b})$. Grid points are generated in the unit hypercube, and transformed to lie within $[\bsym{a}, \bsym{b})$, and function evaluations are computed. For each coordinate, an orthogonal projection is made onto each marginal axis $[a_k, b_k)$ for $k=1,\ldots,s$. Details of the orthogonal projection used in Step 4 are provided in the Appendix. At each abscissa point, the pointwise mean \eqref{eq::PWM} is calculated from the $n^{s-1}$ function evaluations that correspond to the abscissa point (Step 5). An interpolant can then be fitted (such as a cubic spline) and normalised for the grid approximation to the $k^{th}$ marginal. \\

Grid-based methods are flexible in the implementation process. Changes can be made to the general method which can result in similar approximations with less computational effort. For instance, INLA chooses points by only generating important points based off some user-defined criteria which defines the denseness of the grid \cite{rue}. Thus it can save on points which improves computational efficiency. The BISQuE \cite{hewitt} method use sparse grids for approximation to marginal densities, reducing the number of points needed for approximation. Methods described in \cite{pb} generate grid-like structures, such as maximal-rank lattice point sets \cite{disney}, which can better capture the behaviour of the function, improving both accuracy and computational efficiency. However, the number of points still increases exponentially with the dimension, and thus a new approach is required to reduce the number of points further. We believe that the use of an LDS instead of a grid is one such approach, since the number of LDS points do not increase exponentially with dimension.

%%%%%%%%%%%%%%%%%%%%%%%%%%%%%%%%%%%%%%%%%%%%%%

\section{LDS--StM Method and Modifications}
First, we give a brief overview of the LDS--StM method before presenting the modifications. There are two sets of modifications, both of which have corresponding algorithms, the LDS--QA (LDS with quadratic approximation), and the LDS--CX (LDS with correction polynomial of degree X). We present the algorithms and details later this section.

\subsection{Standard LDS method (LDS--StM)}
The LDS--StM was developed in \cite{joshiArticle} as a faster alternative to grid-based methods. In the place of a grid, we instead generate an LDS. Many LDS sequences would do, but we use a Korobov Lattice as described in \eqref{eq:koroPS} and use LatticeBuilder to find appropriate generating constants. Let ${\cal{K}}_{N,s,\alpha}$ be a Korobov lattice with $N$ points in $s$ dimensions, and generating constant $\alpha$. Again, we scale the Korobov lattice points generated from the unit hypercube to the integration region $[\bsym{a}, \bsym{b})$ using the scaling transformation in \eqref{eq::trans}. We present LDS--StM in Algorithm 2. \\

\begin{algorithm} [h]
\caption{LDS--StM Algorithm} \label{alg:LDS}
\begin{algorithmic}
\State 1) Optimise $\pi(\bsym{\theta})$ for mode $\pi^{*}$ and Hessian $H_{\pi}$
\State 2) Construct an appropriate integration region $[\bsym{a}, \bsym{b})$
\State  3) Generate a set of LDS points and transform to $\{\bsym{\theta}_{(j)}\}_{j=1}^{N}$ over $[\bsym{a}, \bsym{b})$ \\ \hspace{3mm} and evaluate $\pi(\bsym{\theta}_{(j)})$ for $j=1,\ldots,N$
\For {$k = 1, \ldots, s$}
\State 4) Orthogonally project points $(\bsym{\theta}_{(j)}, \pi(\bsym{\theta}_{(j)}))$ for $j=1,\ldots,N$ onto \\ \hspace{9mm}  $[a_k, b_k)$
\State 5) Fit least squares polynomial through the orthogonally projected \\ \hspace{9mm} points $(\theta_{k, (j)}, \pi(\bsym{\theta}_{(j)}))$
\State 6) Normalise for $\tilde{\pi}_k(\theta_k)$
\EndFor
\end{algorithmic}
\end{algorithm}

Note that in the LDS--StM algorithm, after orthogonal projections of the points onto the marginal axis in Step 4, we see a scatter of ordinates. These ordinates correspond to $N$ equally spaced and unique abscissa points along the $\theta_k$-axis, which is a result of a property of LDS known as fully projection regular (see \cite{lemieux} for more details). Figure 2 shows the orthogonal projections of the function evaluations onto a single dimension. An interpolant such as a cubic spline that could be used in the general grid method would not be appropriate here, as the resulting approximation would be a spline fitted through each point. Instead, a least squares polynomial is used in Step 5 to approximate the shape of the marginal, and normalisation of the polynomial gives the approximation of the marginal $\tilde{\pi}_k(\theta_k)$. \\

The LDS--StM was shown to be more computationally efficient compared with grid-based methods \cite{joshiArticle}. Most importantly, convergence theorems were proved to show that the approximations converged to the true marginal distribution as the number of points and polynomial degree increased. However, there were some shortcomings with respect to practical implementation. There was no reasonable way to choose the degree of the polynomial. For a regular unimodal, symmetric (or close to) marginal density, numerical experiments indicated that a polynomial of degree eight was found to be generally sufficient. However, a higher degree polynomial, such as one required for a skewed or multimodal density, would often encounter Runge Phenomenon, where oscillations would form in the tails of the polynomial approximation. Inverted Gramian matrices also tended to be ill-conditioned after computation. In higher dimensions, LDS--StM saved many function evaluation points. However, the number of function evaluations were still too high to be of use for a method such as INLA. \\

In order to improve INLA using LDS, the LDS--StM method needed to be modified such that:
\begin{enumerate}
\item The degree of the polynomial is not problem dependent.
\item The method will work well on any posterior without customisation.
\item It remains a computationally efficient method after modifications to achieve 1 and 2.
\end{enumerate}
We now propose two modifications to the LDS--StM method in order to develop a practical method that can be used in INLA, that satisfies the three criteria above.

\begin{figure}[h]
\begin{center}
\includegraphics[scale = 0.16]{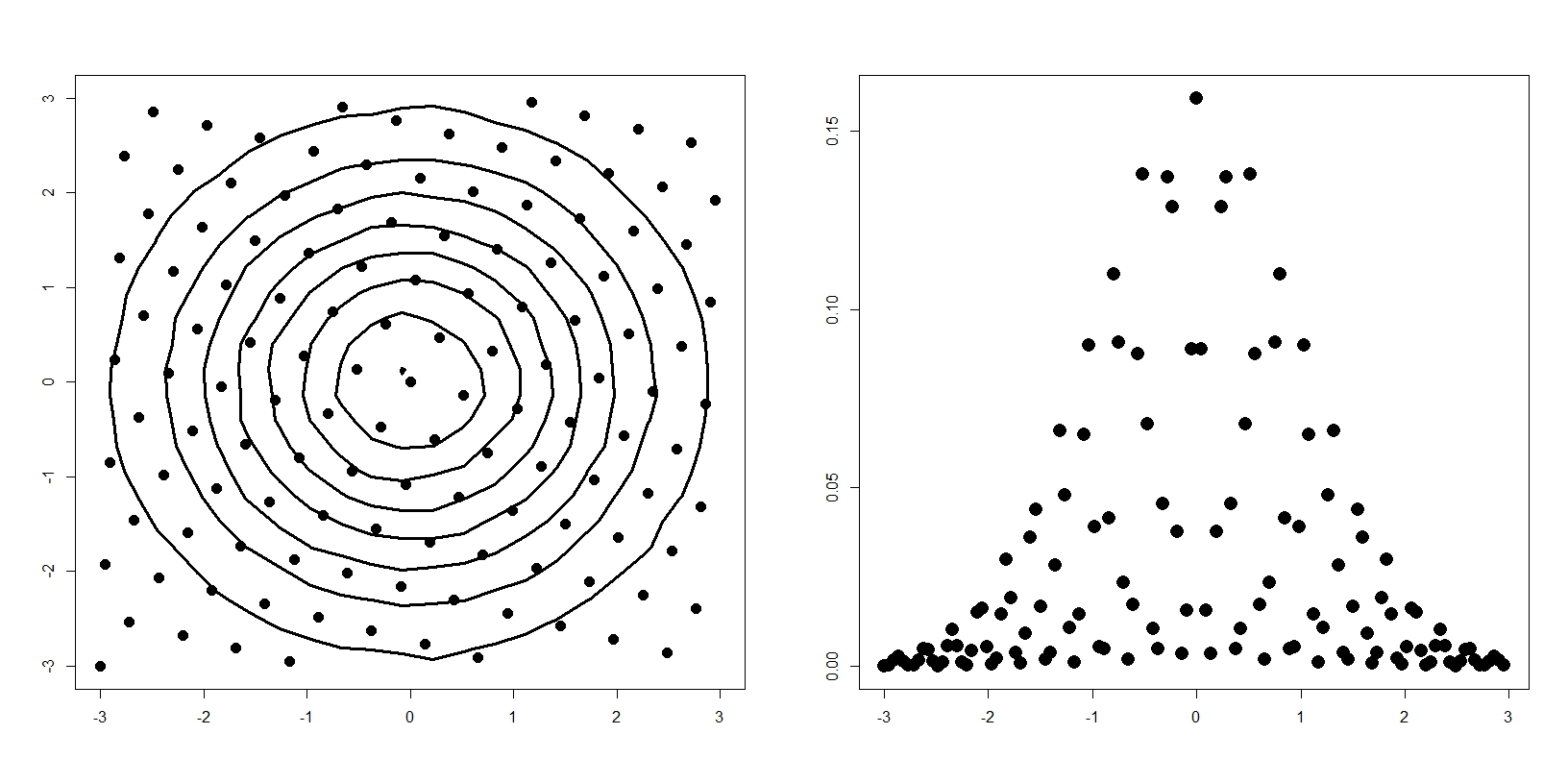}
\caption{\footnotesize{Left: Generating a Korobov point set around a joint posterior. Right: Orthogonally projected points in one dimension.}}
\end{center}
\end{figure}

%%%%%%%%%%%%%%%%%%%%%%%%%%%%%%%%%%%%%%%%%%%%%%%%%%%%%%%%%%
%%%%%%%%%%%%%%%%%%%%%%%%%%%%%%%%%%%%%%%%%%%%%%%%%%%%%%%%%%

\subsection{Initial Modifications - Quadratic Approximation}
Partitioning was used in \cite{joshiArticle} to prove convergence theorems, but not in the LDS--StM algorithm. We decide to include partitioning to aid in the choice of polynomial degree, and it also is useful for the next set of modifications outlined later. For each marginal axis, we construct $n$ partitions. We compute the pointwise mean of each partition (similar to \eqref{eq::PWM}, but calculated at partition intervals rather than unique abscissa points), and fit a polynomial degree of $n-1$ through the pointwise means. \\

The Laplace approximation in \eqref{eq::LaplaceApprox} assumes $\pi(\bsym{\theta}|\bsym{y})$ is almost Gaussian. To aid this assumption, INLA uses variance-stabilising transformations to make each hyperparameter more symmetric \cite{rueReview}. For instance, instead of approximating precisions, INLA approximates $\log$ precisions. Let $\theta_k$ denote a hyperparameter, and let $h$ be the function corresponding to the reparameterisation, then
\begin{equation*}
h(\theta_k) = \theta_{z,k},
\end{equation*}
where $\theta_{z,k}$ denotes the reparameterised hyperparameter. Since these transformations make the hyperparameter close to a Gaussian, this implies we can approximate the marginal using a quadratic polynomial in the $\log$-scale. This also suggests that we should make a minimum of three partitions along the parameter axes. \\

In Algorithm 3, we modify Step 4 in the LDS--StM algorithm to include partitioning and transformations. We also modify part of Step 5 for changes to the quadratic polynomial fit. Since we begin with a reparameterisation of the hyperparameters, Steps 1, 2, and 3 are all performed for the purposes of approximating the marginals $\pi(\theta_{z,k})$. Approximating $\pi(\theta_k)$ requires an inverse transformation from $\theta_{z,k}$ to $\theta_k$.  \\

\begin{algorithm} 
\caption{Quadratic Approximation - LDS--QA} \label{alg:LDSAlgMod}
\begin{algorithmic}
\State 4a) Orthogonally project points $(\bsym{\theta}_{z,(j)}, \pi(\bsym{\theta}_{z,(j)})$ for $j=1,\ldots,N$ onto \\ \hspace{5mm} $[a_k, b_k)$
\State 4b) Create $n$ equally spaced partitions of $[a_k, b_k)$ (the $\theta_{z,k}$ axis). The \\ \hspace{5mm} interval $[\theta_{z,k,u}, \theta_{z,k,u+1})$ for $u=1\ldots,n$ is the $u^{th}$ partition, denoted \\ \hspace{5mm} as $\theta_{z,k,u'}$ with $u' = 1, \ldots, n$
\State 4c) Calculate pointwise means for each partition, where the pointwise  \\ \hspace{5mm} mean of the $u^{th}$ partition is given by $\hat{\pi}(\theta_{z,k,u'})$
\State 5a) Transform pointwise means to $\log(\bsym{\theta}_z)$-scale, and fit a least squares \\ \hspace{5mm} quadratic polynomial through $\log(\hat{\pi}(\theta_{z,k,u})), u = 1,\ldots,n$
\State 5b) Transform polynomial to $\theta_z$-scale and normalise marginal $\tilde{\pi}_k(\theta_{z,k})$, \\ \hspace{5mm} then transform to $\theta$-scale for $\tilde{\pi}_k(\theta_k)$
\end{algorithmic}
\end{algorithm}

We now discuss the LDS--QA algorithm in more detail. After projection of the function evaluation coordinates, we partition the $k^{th}$ marginal axis with $n$ partitions with equal interval lengths. We do not assume that each partition will have the same number of points (for example, having $N = 2^x$ points and $n = 3$ partitions will never yield the same number of points in each partition). However, a Korobov lattice with good generating constant produces $N$ equally-spaced points when projected onto any marginal axis, due to the fully projection regular property. Thus, equally spaced partitions will have a very similar number of points. Let $\nu_u$ denote the number of points in the $u^{th}$ partition. Obtaining the pointwise means is very similar to the grid, though the function evaluations do not project onto a single point, but project onto an interval. The natural abscissa point to take would be to take the midpoint of the interval which is the one we use in practice. Thus, the $u^{th}$ abscissa point would be $0.5 \times (\theta_{z,k,u+1} + \theta_{z,k,u})$. The ordinate for the $u^{th}$ pointwise mean is given by
\begin{equation*} \label{eq::pwmeanMLS}
\hat{\pi}(\theta_{z,k,u'}) = \frac{1}{\nu_u} \sum_{j=1}^{\nu_u} \pi(\bsym{\theta}_{z,u',(j)}),
\end{equation*}
where $\bsym{\theta}_{z,u',(j)}$ are the generated points that lie in the $u^{th}$ partition after being orthogonally projected. Doing this for all $u = 1,\ldots,n$ partitions gives us $n$ pointwise means.\\

We transform the $n$ pointwise means to the $\log(\bsym{\theta}_z)$-scale, and fit the quadratic polynomial via least squares. The quadratic polynomial is given by
\begin{equation} \label{eq::quadratic}
\tilde{\pi}_{k,\log}(\theta_{z,k}) = \tilde{\beta}_{k,0} + \tilde{\beta}_{k,1}\theta_{z,k} + \tilde{\beta}_{k,2}\theta_{z,k}^2,
\end{equation}
where $\tilde{\pi}_{k,\log}(\theta_{z,k})$ is the unnormalised log-approximation to the marginal posterior $\pi_k(\theta_{z,k})$, in which the coefficients $\tilde{\beta}$ are found through a least squares approximation, and $\theta_{z,k} \in [a_k, b_k]$. The quadratic is then transformed back to the $\bsym{\theta}_z$-scale and normalised, which gives the approximation to the reparameterised hyperparameter,
\begin{equation} \label{eq::MLSApproxZ}
\tilde{\pi}_{k}(\theta_{z,k}) = \frac{\exp(\tilde{\beta}_{k,0} + \tilde{\beta}_{k,1}\theta_{z,k} + \tilde{\beta}_{k,2}\theta_{z,k}^2)}{\int_{[a_k, b_k]} \exp(\tilde{\beta}_{k,0} + \tilde{\beta}_{k,1}\theta_{z,k} + \tilde{\beta}_{k,2}\theta_{z,k}^2) d\theta_{z,k}},
\end{equation}
where $\tilde{\pi}_{k}(\theta_{z,k})$ is the approximation of the marginal posterior for $\theta_{z,k}$. For the approximation of the actual hyperparameter, an inverse transformation from $\theta_z$ to $\theta$ is required. \\

\subsection{Second Modification - Polynomial Correction}
From a practical standpoint, the approximation given in \eqref{eq::MLSApproxZ} is far more stable than fitting a polynomial of high degree as performed in the LDS--StM algorithm. However, it is not guaranteed that $\theta_z$'s are exactly Gaussian. Thus, the strategy of fitting a quadratic very much constricts us to a Gaussian estimate after transformation. To account for potential discrepancies in the approximation, we propose another iteration of the algorithm that fits a polynomial with a correction term that can account for skew and multiple modes. This can be done with little extra computational effort. \\

We propose a correction to the quadratic/Gaussian approximation in \eqref{eq::quadratic} by analysing residuals, given by the difference between the polynomial fit and the pointwise means. We name this method the LDS--CX method, where X describes the degree of the correction polynomial. In practice, we typically fit a cubic correction polynomial (LDS--C3) to the quadratic in \eqref{eq::quadratic}, which corrects it for location, scale, and skewness. The process for the LDS--CX algorithm is decribed in Algorithm 4. \\

\begin{algorithm} 
\caption{LDS-CX Algorithm} \label{alg:LDSAlgMod}
\begin{algorithmic}
\State 5a) Transform pointwise means to $\log(\bsym{\theta}_z)$-scale, and fit a least squares \\ \hspace{5mm} quadratic polynomial through $\log(\hat{\pi}(\theta_{z,k,u})), u = 1,\ldots,n$
\State 5a) (i) Obtain residuals found by the difference between $\tilde{\pi}_{\log}(\theta_{z,k})$ \\ \hspace{10mm} evaluated at the $u^{th}$ abscissa point, and its corresponding pointwise \\ \hspace{10mm} mean
\State 5a) (ii) Fit a least squares polynomial (of degree X) through residuals 
\State 5a) (iii) Correct initial quadratic approximation with the polynomial \\ \hspace{12mm} coefficients
\end{algorithmic}
\end{algorithm}

The process changes the LDS--QA algorithm slightly, by adding the polynomial correction process during Step 5a. The correction is summarised in Algorithm 4. After fitting the quadratic polynomial, instead of using this to calculate the posterior marginal, we instead calculate the residuals of the fitted quadratic (the difference between each pointwise mean and the fitted quadratic). After calculating these, we fit a polynomial through these residuals, and use it to update the quadratic. In practice, we generally fit a least squares cubic polynomial, given by
\begin{equation*} \label{eq::cubic}
P_R(\theta_{z,k}) = \beta'_{k,0} + \beta'_{k,1}\theta_{z,k} + \beta'_{k,2}\theta_{z,k}^2 + \beta'_{k,3}\theta_{z,k}^3,
\end{equation*}
where $\theta_{z,k} \in [a_k, b_k]$. Then, the cubic correction update for the unnormalised log-approximation to the marginal posterior $\pi(\theta_{z,k})$ is
\begin{equation} \label{eq::MLSPolyU}
\tilde{\pi}_{\log, P}(\theta_{z,k}) = (\tilde{\beta}_{k,0} - \beta'_{k,0}) + (\tilde{\beta}_{k,1} -  \beta'_{k,1})\theta_{z,k} + (\tilde{\beta}_{k,2} -  \beta'_{k,2})\theta_{z,k}^2 +  \beta'_{k,3}\theta_{z,k}^3.
\end{equation}
From here, Step 5b in Algorithm 3 can be performed to find the final approximations by using the expression in \eqref{eq::MLSPolyU}, substituting into \eqref{eq::MLSApproxZ}, and performing an inverse transformation. \\

As stated previously, we use cubic corrections, mainly to account for any skewness. These approximations almost always outperformed the quadratic approximations, and were much more stable than using the LDS--StM method. In practice, we never used polynomial corrections that were of higher degree than the cubic, with the lone exception being to approximate multimodal densities. We discuss this further in the next section.

%%%%%%%%%%%%%%%%%%%%%%%%%%%%%%%%%%%%%%%%%%%%%%%%%%%%%%%%%%%%%%%%%%%%

\section{Applications and Results}
We have proposed modifications to the LDS--StM method, first by fitting a quadratic as our approximation to the transformed hyperparameters, then by updating the approximation via analysing residuals and correcting the quadratic approximation with a higher degree polynomial (typically a cubic) term. We apply the LDS--QA and LDS--CX method to two examples. The first example is a case study of a spatial analysis of childhood undernutrition in Zambia. The second is a spatio-temporal study of low birth weights in Georgia. Information, details and data for the Zambia example can be found on the R-INLA website (http://www.r-inla.org) and \cite{martino}. Details and links to the data and materials for the Georgia example can be found in \cite{blangiardo}.

\subsection{Child Undernutririon in Zambia}
The Zambia dataset was first introduced by \cite{kandala}. There, the authors used spatial factors to analyse undernutrition among children in the 57 districts (dist) that comprise Zambia. Child undernutrition is measured by the height of a child relative to their age. A $Z$-score is used to determine the stunting of a child, which is defined by
\begin{equation*}
Z_i = \frac{AI_i - MAI}{\sigma}, i = 1\ldots,57
\end{equation*}
where $AI_i$ is the $i^{th}$ child's anthropometric indicator (height relative to age), $MAI$ and $\sigma$ are the median and standard deviation of the referenced population. We assume the scores are conditionally independent Gaussian random variables with unknown mean $\eta_i$ and unknown precision $\tau_1$. \\

Several factors are considered such as age (age), body mass index (bmi) of the child's mother, and several categorical variables including gender, education, mothers employment status and locality. This dataset has been used in \cite{kneib} as an introduction to \emph{BayesX} for the analysis of Bayesian semiparametric regression using MCMC techniques. It was also used to introduce the same idea using INLA. In both studies, they considered the model presented by \cite{kandala}
\begin{equation} \label{eq::ZambiaModel}
\bsym{\eta} = \mu + \bsym{z}_{i}^{T}\bsym{\gamma} + bmi_i \times g_2(dist_i) + g_3(age_i) + g_4(dist_i) + g_5(dist_i).
\end{equation}
Here, $\mu$ is the overall mean, $\bsym{z}_{i}^{T}\bsym{\gamma}$ represent the several categorical covariates $\bsym{z}$ as having a linear effect. Also, $g_5(dist_i)$ is the spatially unstructured component that is i.i.d Gaussian distributed with mean $0$ and unknown precision $\tau_5$, and $g_4(dist_i)$ is the spatially structured component which varies smoothly from district to district. This is modelled as an intrinsic Gauss Markov random field (IGMRF) -- a conditional autoregressive (CAR) prior \cite{besag} -- with unknown precision $\tau_4$. Previous studies believed that $age$ covariate has a non-linear effect, and that the $bmi$ covariate can be used as a weight for the IGMRF $g_2(\cdot)$. Both of these components have unknown precision $\tau_2$ for $bmi$ and $\tau_3$ for $age$. \\

\begin{figure}[h]
\begin{center}
\includegraphics[scale = 0.25]{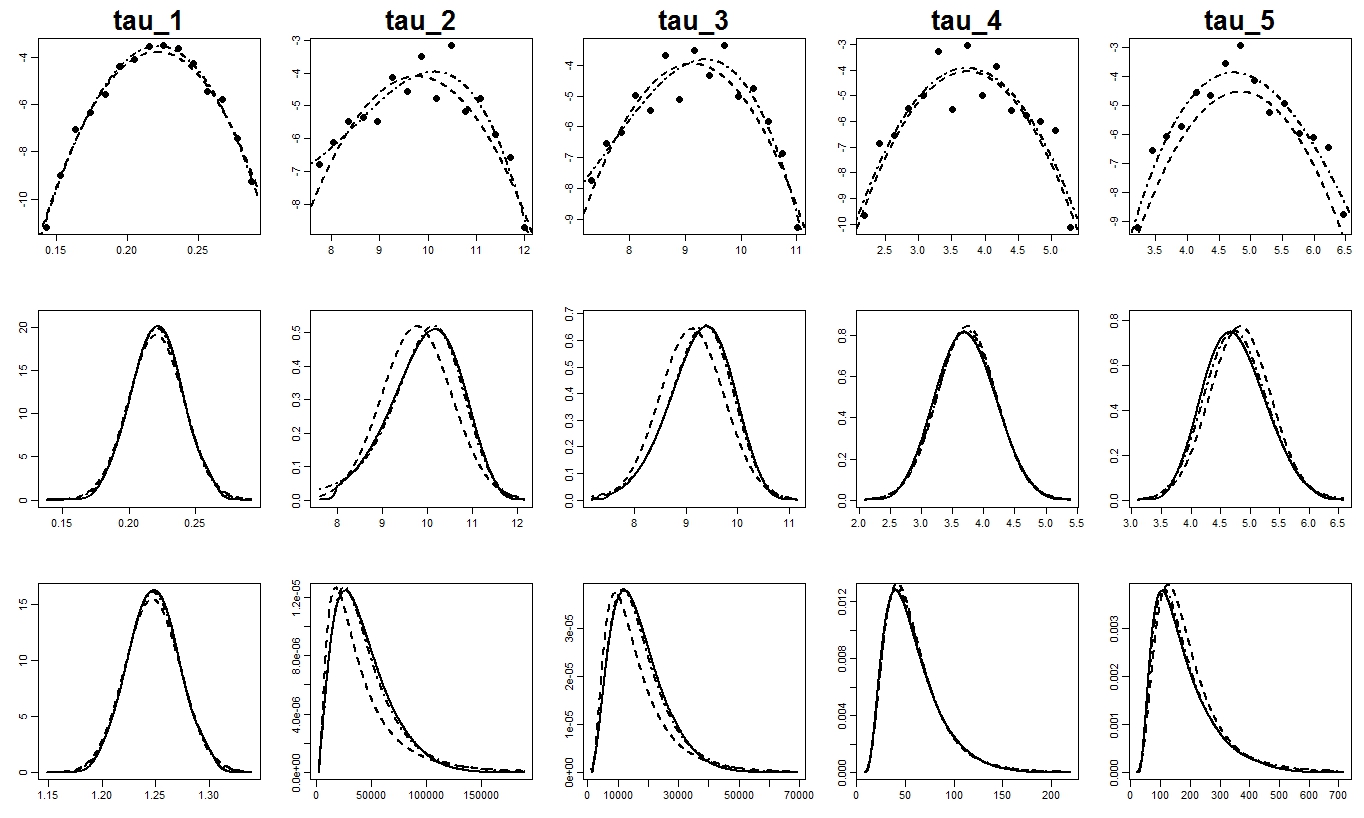}
\caption[Zambia model -- INLA approximations]{\footnotesize{Process of fitting polynomials for the approximations of hyperparameters of the Zambia model. Top row: Fitting a quadratic polynomial (dashed line) and cubic correction (dashed-dotted line) through the pointwise means. Middle row: Approximation of $\bsym{\theta}$ by transforming the polynomials out of the log-scale. The INLA dense grid approximation (bold line) is compared with the initial approximation (dashed line) and the cubic correction (dashed-dotted line). Bottom row: Marginal posterior approximations of $\bsym{\tau}$, showing the cubic correction (dashed-dotted line) providing a more accurate approximation than the initial quadratic approximation (dashed line).}} 
\end{center}
\end{figure}

We will denote the hyperparameters in \eqref{eq::ZambiaModel} as $\bsym{\tau} = \{ \tau_1, \tau_2, \tau_3, \tau_4, \tau_5 \}$. We assign vague Gamma priors for each element in $\bsym{\tau}$. We are interested in estimating the posterior marginal for each hyperparameter. We will first use the grid strategy in INLA before using the methods proposed in Algorithm 3 and Algorithm 4. The reparameterisation INLA uses for the precision hyperparameters (as described in Section 3.2) is a $\log$ transformation. Let $\bsym{\theta} = \{ \log(\tau_1), \log(\tau_2), \log(\tau_3), \log(\tau_4), \log(\tau_5) \}$. To simplify the notation, we will refer to the components of $\bsym{\theta}$ as $\theta_k$ for $k = 1,\ldots,5$.\\ 

For the following results, to get a very precise approximation of the true posterior marginals, we compute the hyperparameters using a customised dense grid in INLA, with a total of around 60000 points. This gives a very good approximation to the true posterior, and we will consider these approximations as the true marginals for the purposes of comparison. \\

\subsubsection{Initial Results}
We apply the steps in the LDS--CX algorithm by updating the quadratic approximation with the coefficients found by fitting a cubic through the residuals. Prior to this, we generate an embedded Korobov Lattice ${\cal{K}}^{*}_{512,5,19}$. Note that $512$ points is less than 116 times the number of points used by INLA's dense grid. For the purposes of this exercise, we use the approximations of the mode and Hessian found by INLA to generate the integration region ($\pm 3$ standard deviations from the mode on each $\bsym{\theta}$-axis). We also use INLA to manually compute the function evaluations for each point in ${\cal{K}}^{*}_{512,5,19}$. After projecting function evaluations onto each axis, we make $15$ equally spaced partitions and find the pointwise means. The log pointwise means and the initial quadratic polynomial fit through those points are shown in Figure 3 (top row) with the points and bold line respectively.\\

A cubic polynomial is fitted to the residuals of the initial fit and the coefficients are used to update the initial fit. The updated fit is shown in Figure 3 as the dashed-dotted line. The initial fit for $\theta_1$ and $\theta_4$ are very good, and as such the cubic correction does not update them. There is a slight skew to $\theta_2$ and $\theta_3$ which cannot be captured by the initial fit. However, the cubic correction captures this skew well. The cubic correction also updates $\theta_5$ slightly, with a small shift in location and skew. The transformations from the approximations in the $\log(\bsym{\theta})$-scale to $\bsym{\theta}$ scale shows how the initial approximation has shifted towards INLA's dense grid approximation in Figure 3 (middle row). It especially highlights how the cubic correction has approximated the two more heavily skewed posterior marginals $\theta_2$ and $\theta_3$. For completeness, we give the results of the inverse transformation from $\bsym{\theta}$ to $\bsym{\tau}$ in Figure 3 (bottom row)). 

\begin{figure}[h]
\begin{center}
\includegraphics[scale = 0.32]{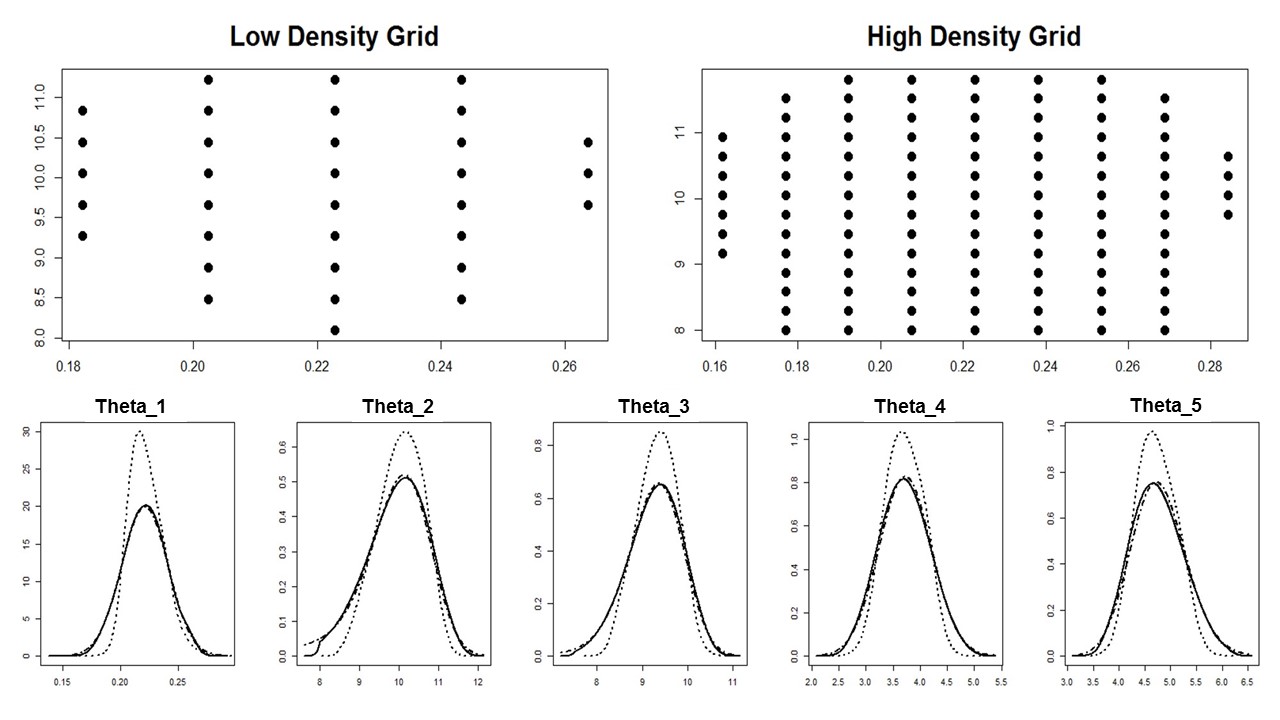}
\caption[INLA's low density grid and high density grid]{\footnotesize{Results of INLA's low density grid. Top row: The grid structures in the dimensions $\theta_1$ and $\theta_2$ for INLA's low density grid and INLA's high density grid. Bottom row: Comparisons with INLA's low density grid and the cubic correction. The grid approximation (dotted line) is poor, despite having roughly five time as many points as the cubic correction (dashed-dotted line).}}
\end{center}
\end{figure}

\subsubsection{Comparisons with INLA Low Density Grid}
We give comparisons of the cubic correction with a grid that is less dense than that given earlier in this section, and has a more comparative number of points to our Korobov Lattice. We used INLA to generate a grid that has $2655$ points, around five times more than that of our Korobov lattice, and far less dense than that of our earlier grid. Figure 4 (top row) shows both the low density grid, and high density grid in two dimensions ($\theta_1$ and $\theta_2$). Whilst this might not look like much difference, the number of points increases exponentially with dimension. The resulting approximations shown in Figure 4 (bottom row) show that the less dense grid failed to give appropriate approximations for all posterior marginals and is outperformed by the cubic correction. 

\subsubsection{Comparisons with NIFA}
NIFA is the current default setting in INLA for the estimation of hyperparameters. Though very fast, there is some room for improvement with respect to accuracy due to the assumed Gaussian structure described in \eqref{eq::NIFA}. We compare the cubic correction with NIFA in Figure 5, and give Kullback-Leibler divergence \cite{kld} and Hellinger distance \cite{hellinger} results in Table 1. \\

\begin{figure}[h]
\begin{center}
\includegraphics[scale = 0.33]{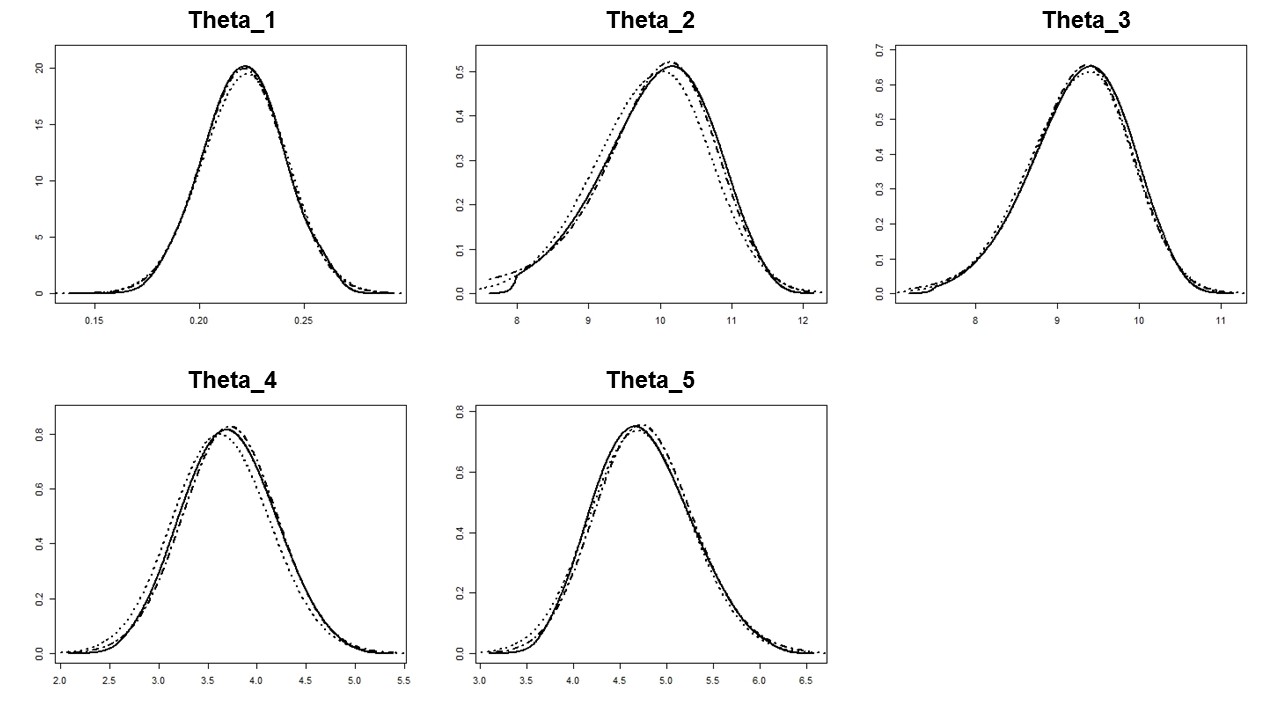}
\caption[Zambia approximations -- NIFA vs cubic correction]{\footnotesize{Zambia: Comparisons between INLA's NIFA (dotted line) and the cubic correction (dashed-dotted line). They are both accurate, though the cubic correction approximates slightly better for all marginals.}}
\end{center}
\end{figure}

The results in Figure 5 show that there is very little difference between the NIFA and the cubic correction. However, the Kullback-Leibler divergence and Hellinger distances in Table 1 shows that for all marginals, the cubic correction did approximate better than NIFA. However, in terms of computational efficiency (points used), NIFA bypasses the need for points altogether, thus approximates much faster than the cubic correction. In our second example, we present why the assumed structure of the NIFA estimate could be a problem, and how a polynomial correction can overcome this problem which can lead to more accurate approximations. \\

\begin{table}[h]
\centering
\caption[Zambia -- Kullback-Leibler divergence and Hellinger distances]{\footnotesize{Zambia: Distance measures comparing INLA's dense grid with both the NIFA and cubic correction methods. The cubic correction gave the more accurate approximations for each hyperparameter marginal according to both the Kullback-Leibler divergence and Hellinger distances, though the differences are very small.\\}}
\begin{tabular}{ |p{2cm}|p{1.5cm}|p{1.5cm}|p{1.5cm}|p{1.5cm}| }
 \hline
  \multicolumn{1}{|c|}{ZAMBIA} &
  \multicolumn{2}{|c|}{K-L.Div} & 
  \multicolumn{2}{|c|}{H.Dist} \\
 \hline
Parameter & NIFA & LDS--C3 & NIFA & LDS--C3 \\
 \hline
$\theta_1$ & 0.00501 & 0.00329 & 0.03961 & 0.03233 \\
$\theta_2$ & 0.01194 & 0.00495 & 0.05664 & 0.04088 \\
$\theta_3$ & 0.00329 & 0.00290 & 0.03058 & 0.02964 \\ 
$\theta_4$ & 0.01199 & 0.00248 & 0.05797 & 0.02655 \\
$\theta_5$ & 0.00787 & 0.00533 & 0.04953 & 0.03967 \\
 \hline
\end{tabular}
\end{table}

\subsection{Low Birth Weight Counts in Georgia}
This example considers the count of new-borns with very low birth weight (less than 2500gm) in the counties of Georgia, USA. The data was collected over ten years from 2000-2010, and used to perform spatio-temporal disease mapping. This particular example has been used by \cite{blangiardo} to illustrate a spatio-temporal Poisson nonparametric approach. This includes interactions between different spatial and temporal components. \\

\begin{figure}[h]
\begin{center}
\includegraphics[scale = 0.33]{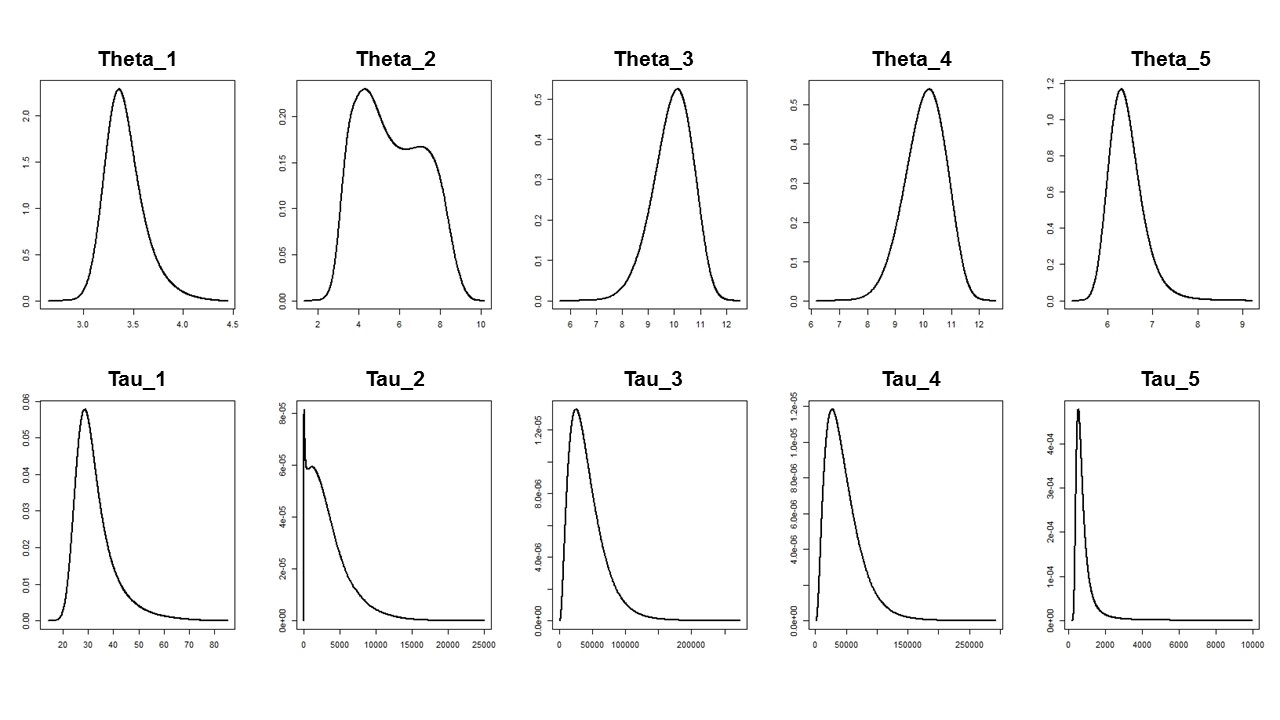}
\caption[Georgia model -- INLA approximations]{\footnotesize{INLA's approximation to the marginal posteriors for each hyperparameter in the Georgia model. With a very dense grid, the second posterior is shown to be multimodal.}}
\end{center}
\end{figure}

Our model is a space-time interaction model, which has the form
\begin{equation*}
\eta_{i,t} = \mu + g_1(county_i) + g_2(county_i) + g_3(year_t) + g_4(year_t) + g_5(area_i \times year_t),
\end{equation*}
where $i = 1, \ldots, 159$ and $t = 1,\ldots,11$. We have $\mu$ as the overall mean, $ g_1(county_i)$ as the spatially structured component, modelled as a conditional autoregressive prior with unknown precision $\tau_1$, and $g_2(county_i)$ as the spatially unstructured component modelled as i.i.d Gaussian with mean $0$ and unknown precision $\tau_2$. Our time components are $g_3(year_t)$, which is modelled as a random walk of order two \cite{lindgrenRue} with unknown precision $\tau_3$, and $g_4(year_t)$ is the unstructured time effect modelled as i.i.d Gaussian with mean $0$ and precision $\tau_4$. We also have an interaction term $g_5(area_i \times year_t)$. We choose an interaction between the unstructured space and unstructured time variables, thus placing no spatial or temporal structure on the interaction, and so modelling this as i.i.d Gaussian with mean $0$, and unknown precision $\tau_5$. Thus we have five hyperparameters $\bsym{\tau} =\{ \tau_1, \tau_2, \tau_3, \tau_4, \tau_5 \}$. Similar to the previous example, we estimate the $\log$ of the precisions, and denote them $\theta_k$ for $k = 1,\ldots,5$. \\

Again, we use a customised, dense grid in INLA (over 100000 points) to compute approximate posteriors that are very accurate. For the purpose of comparison, we will consider these approximations from the dense grid as the true marginals. Figure 6 shows the approximations, and captures a multimodal posterior for $\tau_2$. We present the cubic correction first, before looking at how we can use higher degree polynomials to approximate multimodal posteriors.\\

\begin{figure}[h]
\begin{center}
\includegraphics[scale = 0.33]{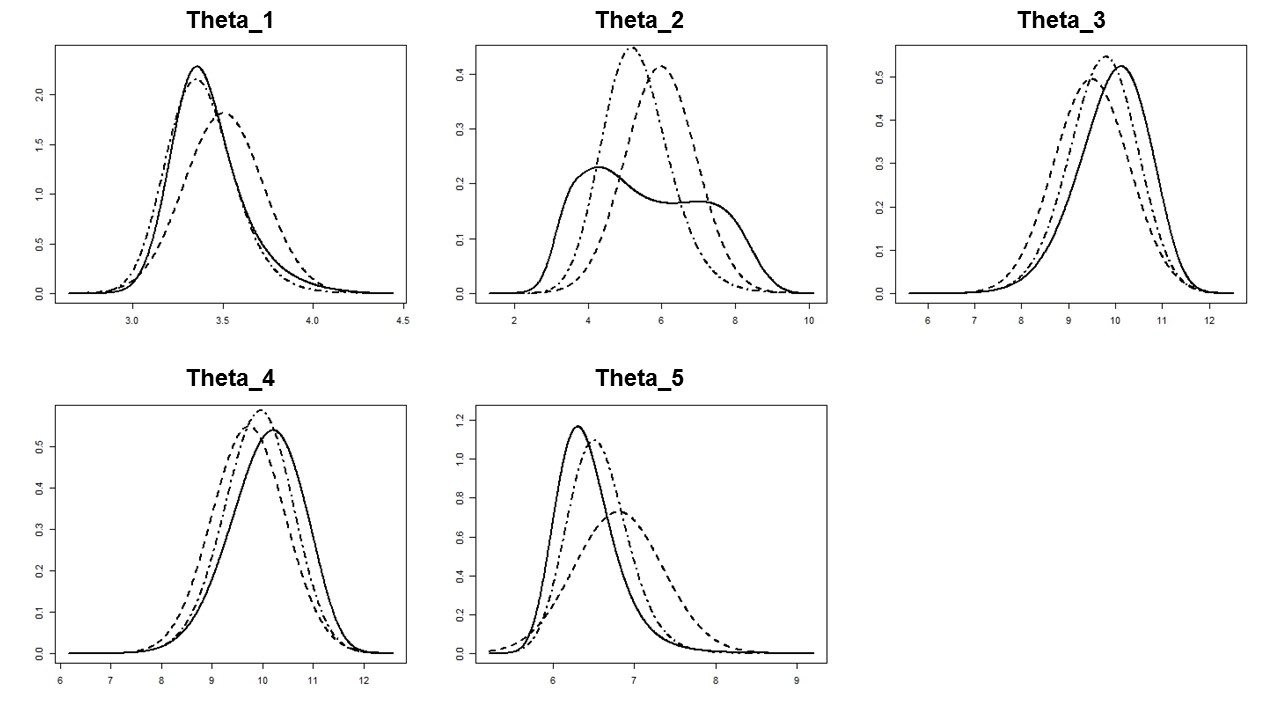}
\caption[Georgia model -- $\bsym{\theta}$ approximations]{\footnotesize{Georgia: Initial Approximation (dashed line) and cubic corrections (dashed-dotted line) for $\bsym{\theta}$, and comparisons with INLA's dense grid (bold line). The cubic correction was needed as the initial fits were not very good. However, we cannot approximate $\theta_2$ well, due to the multimodal nature of the hyperparameter.}}
\end{center}
\end{figure}

\subsubsection{Initial Results}
We take the same approach as we did in the Zambia example by generating an embedded Korobov Lattice ${\cal{K}}_{512, 5, 19}^{*}$, computing the function evaluations and partitioning making 15 equally spaced partitions for each axis. We follow the processes outlined in the LDS--QA and LDS--C3, by fitting an initial quadratic polynomial through the pointwise means, finding the residuals, fitting a cubic polynomial and updating the initial approximation. We display the approximations for the components of $\bsym{\theta}$ in Figure 7. The initial approximations were not good as some of the components of $\bsym{\theta}$ were quite skewed. However, we see that the cubic correction was able to improve the approximations. The approximation of $\theta_2$ was not accurate due to it being multimodal in shape. \\

\subsubsection{Multimodal Hyperparameters - Polynomial Correction}
Up until now we have chosen fit a cubic polynomial to the residuals to correct our initial quadratic approximation for skewness. If a density is unimodal, this is all that is necessary. However, for special cases such as multimodal densities, we can fit a higher degree polynomial correction. Since we fit each hyperparameter independently of the other, this is easily done. We focus solely on the marginal $\theta_2$ and fit both a quartic and quintic correction. \\

\begin{figure}[h]
\begin{center}
\includegraphics[scale = 0.35]{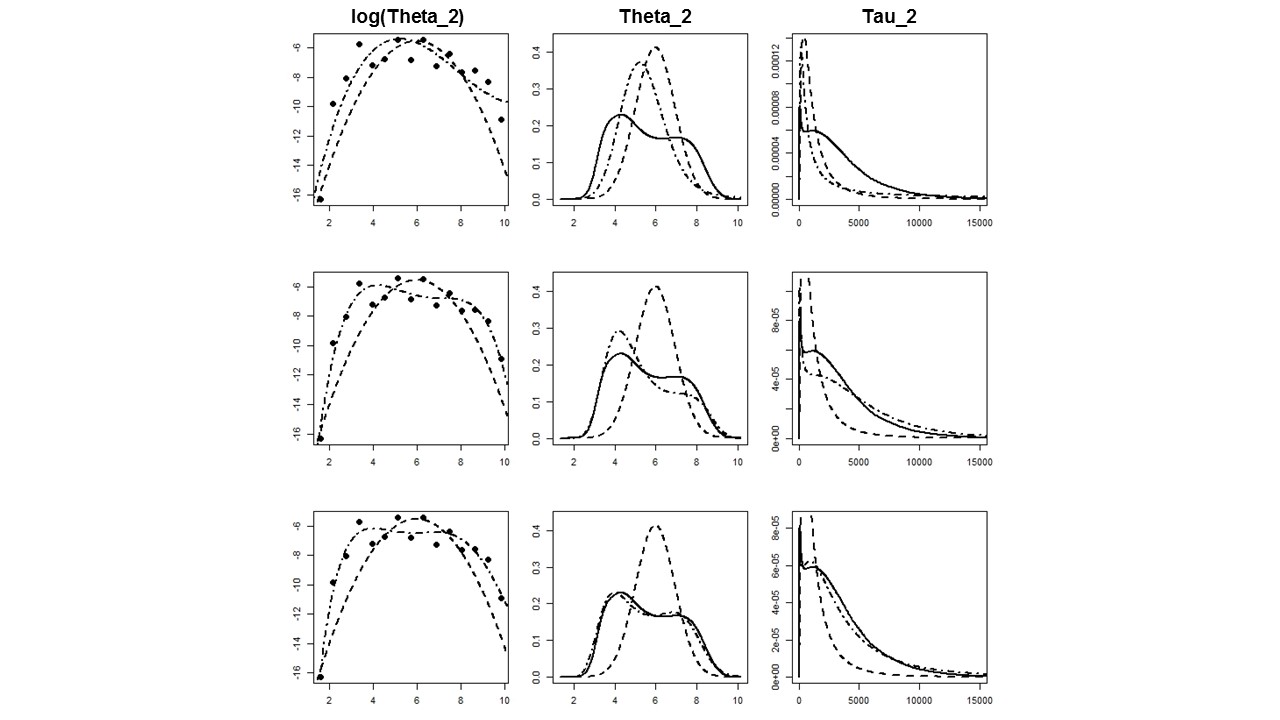}
\caption[Georgia model -- higher order polynomial correction fit]{\footnotesize{Georgia: The process of approximating $\tau_2$ by fitting a cubic correction (top row), a quartic correction (middle row) and quintic correction (bottom row). Note that the first column is fitting the initial quadratic approximation (dashed line), and the polynomial update (dashed-dotted line), whilst the second and third columns are approximating the $\theta_2$ and $\tau_2$ respectively. Going to a higher degree polynomial worked well for approximating the multimodal density.}}
\end{center}
\end{figure}

Figure 8 shows the process of fitting the marginal $\tau_2$ using three different corrections. As shown here and in the initial results, the cubic correction was unable to capture the multimodal shape. The quartic polynomial has up to three turning points, so can detect up to two modes. It did not quite capture both modes. However it was a much better approximation than the cubic. Finally the quintic correction detected both modes and gave an accurate approximation.

\subsubsection{Comparisons with NIFA}

\begin{figure}[h]
\begin{center}
\includegraphics[scale = 0.33]{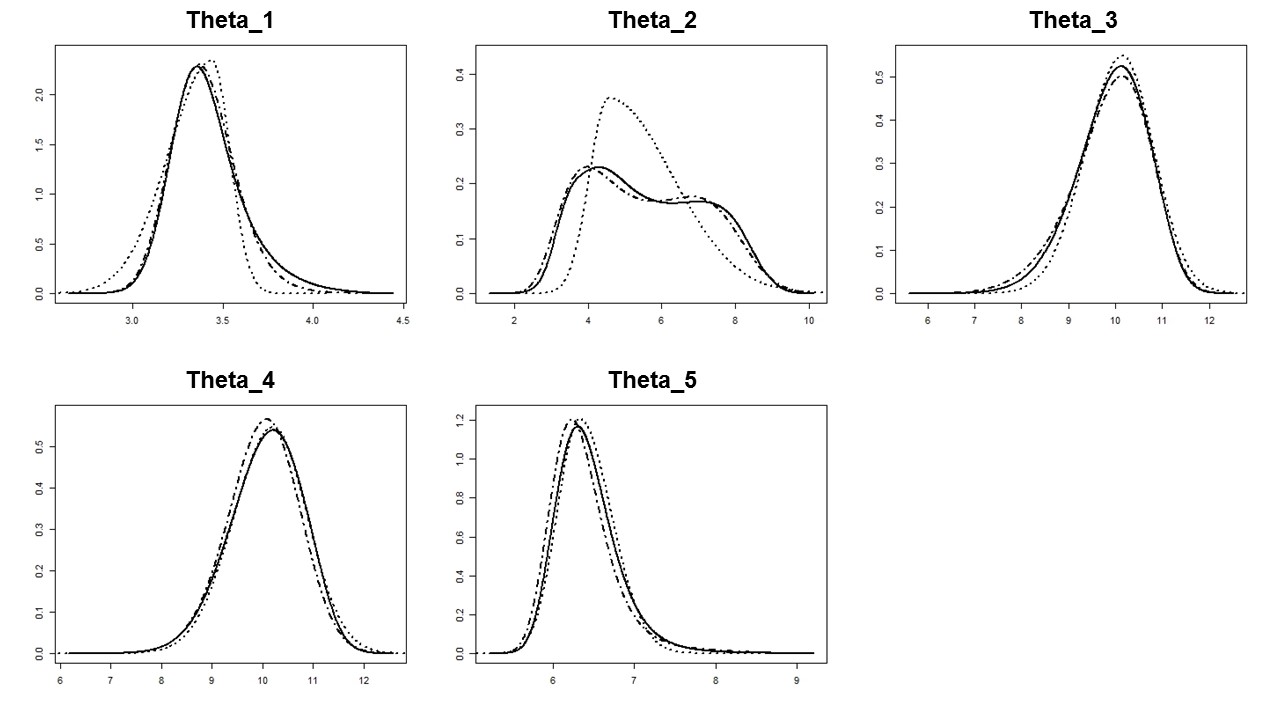}
\caption[Georgia model -- comparisons between NIFA and polynomial correction]{\footnotesize{Georgia: Comparisons between NIFA (dotted line) and polynomial correction (dashed-dotted line). Cubic corrections were used for all hyperparameters, except for $\theta_2$ which has a quintic correction).}}
\end{center}
\end{figure}

We end this section by giving some comparisons with INLA's NIFA method. As discussed previously, the speed and accuracy of NIFA is very good. However, NIFA does assume the marginal is a Gaussian with different standard deviations on each side, hence only being able to give unimodal approximations. We give visual comparisons in Figure 9, and Kullback-Leibler divergence and Hellinger distances between INLA's dense grid (regarded as the true density) and the approximations in Table 2. \\

The polynomial correction we use is cubic for all marginals, except for $\theta_2$, for which we use a quintic correction. As expected, for $\theta_2$, the NIFA approximated this with a unimodal density. The approximation for $\theta_1$ was slightly off for the NIFA too, with the Kullback-Leibler and Hellinger distance being much higher than the cubic correction. The other posterior marginals were well approximated by both methods, and NIFA outperformed the cubic correction for the marginal $\theta_4$. 

\begin{table}[h]
\centering
\caption[Georgia -- Kullback-Leibler divergence and Hellinger distances]{\footnotesize{Georgia: Distance measures comparing INLA's dense grid with both the NIFA and polynomial correction methods. Cubic polynomials were used for all but the second marginal, which used a quintic. The corrections gave the more accurate approximations for each hyperparameter except $\theta_{4}$ according to both the Kullback-Leibler divergence and Hellinger distances.\\}}
\begin{tabular}{ |p{2cm}|p{1.5cm}|p{1.5cm}|p{1.5cm}|p{1.5cm}| }
 \hline
  \multicolumn{1}{|c|}{GEORGIA} &
  \multicolumn{2}{|c|}{K-L.Div} & 
  \multicolumn{2}{|c|}{H.Dist} \\
 \hline
Parameter & NIFA & LDS--CX & NIFA & LDS--CX \\
 \hline
$\theta_1$           & 0.30047 & 0.00923 & 0.17790 & 0.04445 \\
$\theta_2$           & 0.25381 & 0.00601 & 0.22173 & 0.04080 \\
$\theta_3$    & 0.01446 & 0.00250 & 0.05836 & 0.02597 \\
$\theta_4$    & 0.00459 & 0.00671 & 0.03598 & 0.04059 \\
$\theta_5$ & 0.02541 & 0.01550 & 0.06262 & 0.05712 \\
 \hline
\end{tabular}
\end{table}

\section{Discussion and Further Work}

Because of its fully projection regular property, the number of LDS points needed to integrate does not increase rapidly with the number of dimensions. This makes LDS a promising alternative to grid point sets for marginalisation. The LDS--StM method was presented in \cite{joshiArticle} to perform marginalisation of joint densities using LDS point sets. Approximations based on the LDS--StM method converge to the true marginal, and were shown to be more computationally efficient than existing grid--based methods of marginalisation. The motivation for this work was to build upon the LDS--StM method so it can be implemented within INLA to estimate model hyperparameters. To do this, we focussed on improving the interpolation process and simplifying the polynomials used. An initial quadratic/Gaussian approximation is fit to the pointwise means of orthogonally projected function evaluations (LDS--QA method), and another polynomial is used to correct the initial approximation (LDS--CX method). In practice, we typically use a cubic correction polynomial (LDS--C3), but for more complicated marginal shapes, a higher degree polynomial can be used. \\   

The results for the LDS--CX method are very promising. First, the approximations are far more stable than LDS--StM methods, since we do not need to use high degree polynomials in the interpolation process, thus eliminating ill-conditioned matrices and Runge's phenomenon. This makes the method far more practical than LDS--StM, and useable within INLA. Second, the results in Section 4 show that the approximations using LDS--CX outperform INLA's grid by several orders of magnitude, with respect to computational efficiency. In higher dimension, we can obtain as accurate approximations than INLA's grids with far less points. Also, for any given number of points, we can obtain better approximations by using LDS and our algorithms within INLA, than using INLA's grid. Lastly, the LDS--CX method has the potential for estimating multimodal marginals by increasing the degree of the correction polynomial. This is not possible with INLA's NIFA approximation as it assumes the marginal posterior has a unimodal, ``half" Gaussian shape, with different standard deviations either side of the mode (as shown in Equation \eqref{eq::NIFA}). Further investigation is needed to develop a diagnostic which allows the user to detect multimodal marginal hyperparameters, and determines which correction polynomial to use. \\

This paper was motivated by the efficient approximation of hyperparameters using LDS. However, there is also the potential for computing efficient approximations for the latent parameters too. To estimate latent parameters of an LGM, INLA re-uses the points generated via grid (or otherwise) in the numerical integration process. Given that the dimension of latent fields are typically large, this will require a lot of computing power if we require many points to estimate the hyperparameters. Recall from Section 2.1, that an extensible lattice can be generated to give the user the ability to add more points without discarding all the old points. We can also take points away in a similar way without affecting the low discrepancy structure of the point set (see Figure 1). For example, let ${\cal{K}}_{2^x, s,\alpha}^{*}$ be an extensible Korobov lattice with $2^x$ points in $s$ dimensions. Expressing the pointset as a matrix form, keeping the first row and taking out every second row after will give ${\cal{K}}_{2^{x-1}, s,\alpha}^{*}$, and keeping the first row and keeping every fourth row will give ${\cal{K}}_{2^{x-2}, s,\alpha}^{*}$, and so on. This subset of points can be used to approximate latent posterior marginals, as described in \eqref{eq::latentMarginals}. Recent developments by the INLA team allow any user-defined pointset for the estimation of latent parameters. Further work is needed here to test the accuracy and computational efficiency of latent parameter estimates using LDS and extensible Korobov lattices. 

\newpage

\section*{Appendix - Orthogonal Projections}
Let  $f(\boldsymbol{x})$ be an $s$ dimensional function that is evaluated at $N$ unique points $\boldsymbol{x}_{N}={\boldsymbol{x}_{(1)},\ldots, \boldsymbol{x}_{(N)}}$, where each $\boldsymbol{x}_{(i)} \in[\bsym{a}, \bsym{b}),\, i=1,\ldots,N,$ is an $s-tuple$ $\boldsymbol{x}_{(i)}=(x_{1,i},\ldots,x_{s,i}).$ These points along with the function evaluations, that is $(\boldsymbol{x}_{(i)}, f(\boldsymbol{x}_{(i)}))$ are $(s+1)-tuples$ that can be expressed in matrix form as
\begin{eqnarray*}
\boldsymbol{\Psi} = 
\begin{bmatrix}
    x_{1,1}  & \dots  & x_{s,1} & f(\boldsymbol{x}_{(1)})   \\
    x_{1,2}  & \dots  & x_{s,2} & f(\boldsymbol{x}_{(2)})  \\
    \vdots  & \ddots & \vdots & \vdots\\
    x_{1,N}  & \dots  & x_{s,N} & f(\boldsymbol{x}_{(N)})  \\
\end{bmatrix}
\end{eqnarray*}

To estimate the $k^{th}$ marginal $f_k(x_k)$, we orthogonally project $(\boldsymbol{x}_{(i)}, f(\boldsymbol{x}_{(i)}))$ for $i = 1,\ldots,N$ on the $ k^{th}$ marginal axis to obtain $\psi_{k} = P_{k}\boldsymbol{\Psi}$, 
\begin{eqnarray*}
\psi_k = 
\begin{bmatrix}
    x_{k,1} & f(\boldsymbol{x}_{(1)})   \\
    x_{k,2} & f(\boldsymbol{x}_{(2)})  \\
    \vdots   & \vdots\\
    x_{k,N} & f(\boldsymbol{x}_{(N)}) \\
\end{bmatrix},
\end{eqnarray*}
where $P_{k} = A_k(A_{k}^\top A_k)^{-1}A_{k}^\top$ is a projection matrix and $A_k$ has size $(s+1)\times 2$, and is a unit basis vector for $\mathbb{R}^{2}$ with the $k^{th}$ entry in the first column and the $(s+1)^{th}$ entry in the second column  as one, all the remaining entries are zeros. \\ 

For example, if $s=3$ and $k=2$ then,
\begin{eqnarray*}
\boldsymbol{\Psi}= 
\begin{bmatrix}
    x_{1,1}  & x_{2,1} & x_{3,1} & f(\boldsymbol{x}_{(1)})   \\
    x_{2,1}  &  x_{2,2}  & x_{3,2} & f(\boldsymbol{x}_{(2)})  \\
    \vdots  & \ddots & \vdots & \vdots\\
    x_{1, N}  &  x_{2,N}  & x_{3,N} & f(\boldsymbol{x}_{(N)})  \\
\end{bmatrix}, \;
A_2 =  
\begin{bmatrix}
0 & 0\\
1 & 0 \\
0 & 0 \\
0 & 1\\
\end{bmatrix}, \;
P_2 = 
\begin{bmatrix}
0 & 0 & 0 & 0\\
0 & 1 & 0 & 0\\
0 & 0 & 0 & 0\\
0 & 0 & 0 & 1\\
\end{bmatrix}
\end{eqnarray*}
and 
\begin{eqnarray*}
P_2 \bsym{\Psi} = 
\begin{bmatrix}
   0 &  x_{2, 1}  & 0 & f(\boldsymbol{x}_{(1)})   \\
   0 &  x_{2,2} & 0 &f(\boldsymbol{x}_{(2)})  \\
   \vdots & \vdots & \vdots   & \vdots\\
    0 & x_{2,N} & 0 & f(\boldsymbol{x}_{(N)}) \\
\end{bmatrix}, \mbox{ ignoring rows with zeros }
\begin{bmatrix}
    x_{2,1} & f(\boldsymbol{x}_{(1)})   \\
    x_{2,2} & f(\boldsymbol{x}_{(2)})  \\
     \vdots  & \vdots\\
    x_{2,N} & f(\boldsymbol{x}_{(N)}) \\
\end{bmatrix} = \psi_2.
\end{eqnarray*}

%%%%%%%%%%%%%%%%%%%%%%%%%%%%%%%%%%%%%%%%%%%%%%

\newpage
%\section*{References}

\end{document}